\documentclass[preprint,showpacs,preprintnumbers,amsmath,amssymb]{revtex4-1}

\usepackage{graphicx}% Include figure files
\usepackage{dcolumn}% Align table columns on decimal point
\usepackage{bm}% bold math
\begin{document}

\title{A dynamical model of surrogate reactions}

\author{Y.~Aritomo, S.~Chiba and K.~Nishio}

\affiliation{Advanced Science Research Center, Japan Atomic Energy Agency,
Tokai, Ibaraki, 319-1195, Japan}%
%\affiliation{$^{2}$ Flerov Laboratory of Nuclear Reactions, JINR, Dubna, Russia}%
%\date{\today}% It is always \today, today,
             %  but any date may be explicitly specified

\begin{abstract}

A new dynamical model is developed to describe the whole process of surrogate
reactions; transfer of several nucleons
at an initial stage, thermal equilibration of residues leading
to washing out of shell effects and decay of populated compound nuclei
are treated in a unified framework.  Multi-dimensional
Langevin equations are employed to describe time-evolution of
collective coordinates
with a time-dependent potential energy surface
corresponding to different stages of surrogate reactions.  The new model
is capable of calculating spin distributions of the compound nuclei,
one of the most important quantity in the surrogate technique.
Furthermore, various
observables of surrogate reactions can be calculated, e.g.,
energy and angular distribution of ejectile, and
mass distributions of fission fragments.
These features are important to assess
validity of the proposed model itself,
to understand mechanisms of the surrogate reactions and
to determine unknown parameters of the model.
It is found that spin distributions of compound nuclei
produced in $^{18}$O+$^{238}$U $\rightarrow ^{16}$O+$^{240*}$U and
$^{18}$O+$^{236}$U $\rightarrow ^{16}$O+$^{238*}$U
reactions are equivalent and much less than 10$\hbar$, therefore satisfy
conditions proposed by Chiba and Iwamoto (PRC 81, 044604(2010))
if they are used as a pair in the surrogate ratio method.
\end{abstract}


%\documentclass{elsart}
%\usepackage{natbib}
%\usepackage{graphicx}
%%\renewcommand{\baselinestretch}{1.65}
%%\usepackage[figuresright]{rotating}
%\begin{document}
%\runauthor{Y.~Aritomo}
%\begin{frontmatter}
%\title{Analysis of dynamical process using mass distribution of fission fragments in heavy-ion reactions }

%\begin{document}

\maketitle

% main text

%%%%%%%%%%%%%%%%%%%%%%%%%%%%%%%%%%%%%%%%%%%%
%% MAINMATTER
%%%%%%%%%%%%%%%%%%%%%%%%%%%%%%%%%%%%%%%%%%%%

\section{Introduction}

%----(2010.06.24.)--------------------------------------------------------------------------------------

Neutron-induced cross section data of unstable nuclei are systematically
required to design next-generation nuclear facilities such as high burn-up
fast breeder reactors or accelerator-driven systems for transmutation of nuclear wastes \cite{iaea08}.
Such data are also important to understand origin of elements, namely,
the s- and r-process nucleosynthesis (see, e.g., Refs.~\cite{arnett,pagel}).
However, it is not usually possible to  measure these cross sections directly
by using neutrons due to difficulty in preparing samples.
It is well known that the compound
reaction process is the dominant mechanism in the energy region
of our interest.
Therefore, various experimental
methods to measure
the ``direct" neutron capture components are not be applicable
to determine them.
Instead, other methods are needed, and one of the promising
method is the surrogate
reaction approach
 \cite{cram70a,cram70b,back74a,back74b,brit79,youn03a,youn03b,peti04,plet05,boye06,burk06,esch06,lyle07,
 fors07,jura08,lesh09,allm09,hatarik10,goldblum10,escher10,scielzo10,kessedjian10}.
In this method, (multi) nucleon transfer
reactions with an experimentally accessible combination of projectile and
target are employed to create the same compound nucleus as the desired
neutron reaction, and decay branching ratios to specific channels, normally capture and/or fission,
are determined. However, branching ratios are sensitive to the
spin and parity of the compound state, while the spin-parity ($J^\pi$)
distributions of
populated nuclei are probably different for the neutron-induced
and surrogate reactions.  It is well known that if the spin is different even
just 1 unit, the capture branching ratio is totally different for the energy
region of our interest.  Therefore, validity of the surrogate method depends
on how the difference of the spin-parity
distributions is comprehended and compensated properly.

Recently, the surrogate ratio method (SRM) is discussed by Chiba and Iwamoto.
It was found that SRM works to a certain accuracy if
(1) there exist two surrogate reactions whose spin-parity
distributions of decaying nuclei are equivalent, (2)
difference of representative spin values
between the neutron-induced and surrogate
reactions is not much larger than 10 $\hbar$, under a condition that
(3) weak Weisskopf-Ewing condition, namely, $J^\pi$-by-$J^\pi$
convergence of the branching ratio, is realized \cite{chiba10}.
They form a set of sufficient conditions for the SRM to work.
It is important to notice that the
$J^\pi$ distribution may be even different for
the neutron-induced and surrogate reactions if these conditions are fulfilled.

Discovery of the above conditions
is a great advancement for the whole surrogate technique.
Therefore we need further investigation to verify
that the above conditions, especially (1) and (2) which were just
assumed in Ref.~\cite{chiba10},
are really satisfied in certain surrogate reactions.
It implies that mechanisms of the surrogate reactions to be understood well.
For that aim,
it is indispensable to establish
a theoretical model to describe the whole process
of the surrogate reactions, namely,
nucleon transfer and decay of populated compound nucleus.

In this work, we propose a first version of our model to describe the surrogate reactions
based on a theory proposed originally by Zagrebaev and Greiner \cite{zagr05}.
This model, called a unified model, can treat the whole reaction processes in heavy- and
superheavy-mass regions, which has been applied to several types of reactions \cite{zagr05,zagr07a,zagrt,arit09}.
The name of unified model implies an unified dynamical approach and unified multidimensional potential energy.
Time-evolution of the system is described by a trajectory calculation
on the time-dependent unified potential energy surface using
the Langevin equation.
Then, we
treat a two neutron transfer reaction;
$^{18}$O+$^{238}$U $\rightarrow ^{16}$O+$^{240}$U, which is planned to be employed
at Japan Atomic Energy Agency (JAEA) as an example of application of the new model to the surrogate reaction.
By using the new model, we can obtain various quantities which can be compared with experimental data directly
and we can evaluate our theory and determine unknown parameters in the model.

The purpose of this paper is to explain the new model, calculate
a fission fragment mass distribution (FFMD) for a reaction
$^{18}$O+$^{238}$U to calibrate
the model parameter, and calculate
spin distributions of compound nuclei for the
$^{18}$O+$^{A}$U $\rightarrow ^{16}$O+$^{A+2*}$U system,
 where $A=$236 and 238,
to see if the conditions proposed by Chiba and Iwamoto are satisfied or not.
In section~2, we explain our theoretical framework.
The calculation results  are presented in section~3.
In section~4, we present a summary of this study and further discussion.

%----------------------<
%%%%%%%%%%%%%%%%%%%%%%%%%%%%%%%%%%%%%%%%%%%%%%%%%%%%%%%%%%%%%%%%%%%%%%%%%%%%%%
\section{Dynamical model} %2009.03.02

\subsection{Overview of the model}

The surrogate
reactions consist of 2 stages; an initial nucleon transfer process
and decay of populated compound nuclei, which have quite different
nature to each other.
The Hauser-Freshbach (HF) theory \cite{haus52,freb96} has been applied to
describe the latter part of the surrogate reactions.
By the HF theory, we are able to calculate decay branching ratios to
specific channels (capture or fission) with arbitrary
spin-parity distributions of compound states.  However,
we cannot predict the spin distribution produced by the initial stage
of the surrogate reaction nor the FFMD with the HF theory.
For that reason, we need to describe the whole reaction process consistently,
i.e.,
starting from the transfer of several nucleons
and the decay of the compound nucleus leading to fission successively.
Here, we employ a dynamical model, the unified model, to the surrogate reaction.

The unified model was proposed by Zagrebaev and Greiner and was applied to
several types of reactions \cite{zagr05,zagr07a,zagrt,arit09} induced by heavy ions.
An unified dynamical approach and unified multidimensional potential energy are
employed in this model, which are the origin of the name of this theory.
To apply this model to
the surrogate reaction, we extend this model and introduce new procedures.
As explained above, the surrogate reactions consist of two processes;
the transfer reaction process between a two-body system and
decay of the populated compound nuclei
(one-body system),
for which the mass of total system is different very much to each
other.
Therefore, it is indispensable to connect such different systems
to treat surrogate reactions consistently.
Evolution of the mass-asymmetry
parameter is described by multi-dimensional
Langevin equations without (with) the
inertia parameter before (after) the window of the colliding nuclei opens sufficiently.
We modify the original unified model \cite{zagr05} also
to take account of temperature
dependence of the shell correction energy of the potential energy surface.

Firstly, we treat the transfer reaction process within the framework of the unified model.
Then, we treat the decay of the compound nuclei with an initial
condition populated by the former reaction process.
We perform a trajectory calculation on a time-dependent
potential energy surface
corresponding to different stages of the surrogate
reaction.
A dynamical calculation is carried out
in terms of the multi-dimensional Langevin equation based on the
fluctuation-dissipation theorem.
By this procedure, we can describe trajectories
moving on the potential energy surface including
the nucleon transfer reaction.
%The evolution of the heavy nuclear system can be traced starting from an
%infinite distance between the projectile and target to the end of each process.

We consider in this paper a
two neutron transfer reaction
$^{18}$O+$^{238}$U $\rightarrow ^{16}$O+$^{240*}$U
 as an example.
In the transfer reaction process, we use the potential energy of $^{256}$Fm
as the total system.
After the production of the compound nucleus by the transfer reaction, we then
treat the decay
of the compound nucleus $^{240}$U.
In other words, the potential energy
surface switches from that of $^{256}$Fm to that of $^{240}$U
as the surrogate reactions proceed from the transfer
process to the decay stage.

\begin{figure}
\centerline{
\includegraphics[height=.40\textheight]{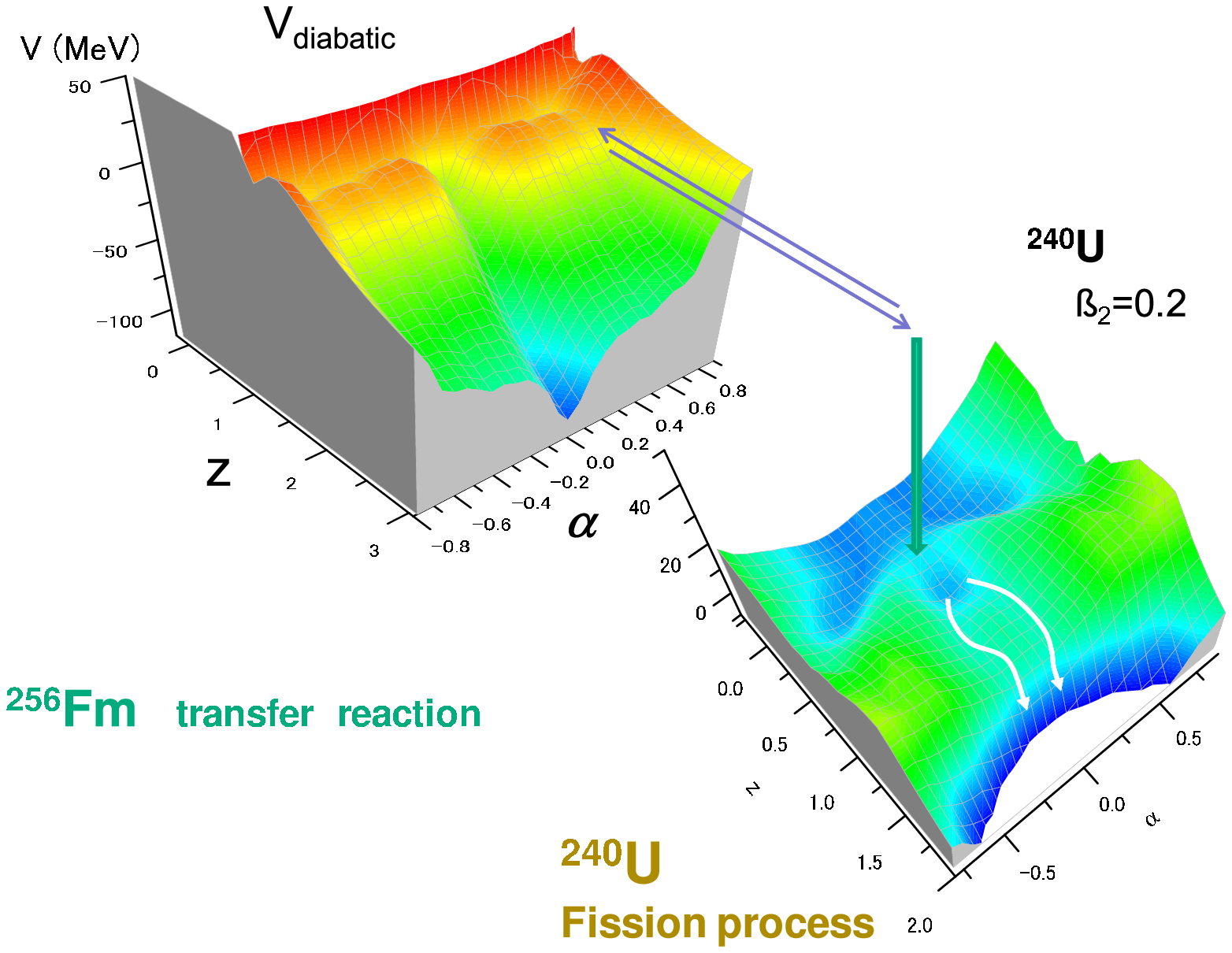}}
  \caption{(Color online) Schematic picture of the calculation. For the surrogate reaction
$^{18}$O+$^{238}$U $\rightarrow ^{16}$O+$^{240}$U, the potential energy surfaces from the
transfer reaction to the decay process of compound nuclei are presented.
The transfer reaction is shown in the left panel which is the diabatic potential
energy surface of $^{256}$Fm in the $z-\alpha (\delta=0)$ coordinate space.
The decay process of compound nucleus is presented in the right panel, which is the
adiabatic potential
energy surface of $^{240}$U with $\delta=0.2$ ($\beta_{2} \sim 0.2$) .}
\label{1-3dim1}
\end{figure}

A schematic picture of our model is presented in Fig.~\ref{1-3dim1}.
It shows the potential energy surfaces used in the trajectory calculation from the
transfer process to the decay of compound nuclei in the reaction
$^{18}$O+$^{238}$U $\rightarrow ^{16}$O+$^{240}$U.
The transfer reaction process is presented in the upper left panel of
Fig.~1 which shows a diabatic potential
energy surface of $^{256}$Fm
in the $z$-$\alpha$ $(\delta=0)$ coordinate space.
Meanings of these parameters and terms are explained
in the next subsection.
The thin arrows correspond to the entrance and the exit channels of two-neutron transfer process.
In the calculation, it starts from an infinite distance between the projectile and target, where
actually the distance of 30 fm is used.
Then, the calculation stops when the trajectory reaches at the distance of 25 fm between the both fragments.
We select the two nucleon transfer among the all events; we select events in which
the mass asymmetry parameter $\alpha$
changes from 0.859 corresponding to $^{18}$O+$^{238}$U to 0.875 corresponding to $^{16}$O+$^{240}$U.

Next, the decay process of compound nucleus is presented in the lower right
panel of Fig.~1, which is a potential
energy surface of $^{240}$U with $\delta=0.2$ ($\beta_{2}\sim 0.2$).
The white lines denote mean fission paths.
In the decay process, we start the trajectory calculation with the initial condition
obtained in the transfer reaction process.
As the initial condition, we use the deformation, momentum, angular momentum and excitation energy of
the compound nucleus.
The quantities which we can obtain with this calculation are the angular and energy distributions of
ejectile, mass and total kinetic energy distributions of fission fragments and neutron multiplicities,
and so on.
Such quantities can be compared with experimental data directly, which allows us to
determine unknown parameter in the model. In this work,
we discuss a mass distribution of fission fragments
and determine a value of one of the most uncertain parameter, the
sliding friction (see below).
In this way, we can evaluate and improve our model step-by-step
by comparing model predictions with experimental data.
More details of our model will be explained below.

%%%%%%%%%%%%%%%%%%%%%%%%%%%%%%%%%%%%%%%%%%%%%%%%%%%%%%%%%%%%%%%%%%%%%%%%%%%%%%
\subsection{Potential energy surface}

The initial stage of the surrogate reaction consists of 2 parts;
1) a fast diabatic part in which the reaction proceeds too fast for nucleons to
reconfigure their single-particle states so the system goes through
the ground-state configurations of the target and projectile, and
2) the system relaxes to the ground-state of the total composite
system, which changes the potential energy surface to an adiabatic one.
Therefore, we take into account time evolution of the
potential energies from the diabatic one
$V_{diab}(q)$ to the adiabatic one $V_{adiab}(q)$, here $q$
denotes a set of collective coordinates representing the nuclear deformation.
The diabatic potential is calculated by a folding procedure with
effective nucleon-nucleon interaction \cite{zagr05,zagr07a,zagr07}, which is
shown in the upper left part of Fig.~\ref{1-3dim1}.  We can see a ``potential wall'' in
the overlap region of the colliding system, which corresponds to a
hard-core representing incompressibility of nuclear matter.
On the other hand, the adiabatic potential energy of the system is calculated
using an extended two-center shell model \cite{zagr07}.
We then connect the diabatic and adiabatic potentials with a
time-dependent weighting function as follows;
\begin{eqnarray}
&&V=V_{diab}(q)f(t)+V_{adiab}(q)[1-f(t)], \nonumber \\
&&f(t)= \exp \left (-\frac{t}{\tau} \right ).
\end{eqnarray}
Here, $t$ is the time of interaction and $f(t)$ is the
weighting function with the
relaxation time $\tau$.
We use a relaxation time $\tau=10^{-21}$ sec, which was suggested
in references \cite{bert78}.  It is empirically known that calculated
results do not depend noticeably on the relaxation time.

% epsilon

As the coordinates to express nuclear deformation, we use the two-center
parametrization \cite{maru72,sato78} and employ three parameters as follows:
$z_{0}$ (distance between centers of two potentials), $\delta$
(deformation of fragments), and $\alpha$ (mass asymmetry of the
colliding nuclei); $\alpha=(A_{1}-A_{2})/(A_{1}+A_{2})$, where
$A_{1}$ and $A_{2}$ denote the mass numbers of the target and the
projectile, respectively \cite{arit04}.  Later on, $A_1$ and $A_2$ are used
to denote
mass numbers of two fission fragments.
The parameter $\delta$ is defined as $\delta=3(a-b)/(2a+b)$, where
$a$ and $b$ denote the half length of
the axes of ellipse in the $z_{0}$ and $\rho$ direction, respectively as
expressed in Fig.~1 in reference \cite{maru72}.
We assume that each fragment has the same deformations as a first step.
%Fragment deformation $\delta$ is related to more-familiar deformation parameter $\beta_{2}$,
%
%\begin{equation}
%\beta_{2}=\frac{\delta}{\sqrt{\frac{5}{16\pi}}(3-\delta)},
%\end{equation}
%
%here $\delta < 1.5$, because of $a > 0$ and $b > 0 $.
Furthermore, we use scaling to save computation time and employ a coordinate $z$
defined as $z=z_{0}/(R_{CN}B)$,
where $R_{CN}$ denotes the radius of the spherical
compound nucleus.
The parameter $B$ is defined as $B=(3+\delta)/(3-2\delta)$.

In the
two-center parametrization, the neck parameter is denoted by $\epsilon$ and is
known to be different
in the entrance and exit channels \cite{zagr07}.
Therefore, we employ $\epsilon=1$ for the entrance channel and $\epsilon=0.35$ for the exit
channel to describe a realistic nuclear shape.
We introduce a time-dependent potential energy surface in terms of $\epsilon$
using a relaxation time for $\epsilon$ of $\tau_{\epsilon}=10^{-20}$ sec \cite{karp07}, as follows;
\begin{eqnarray}
&&V_{adiab}=V_{adiab}(q,\epsilon=1)f_{\epsilon}(t)+V_{adiab}(q,\epsilon=0.35)[1-f_{\epsilon}(t)], \nonumber \\
&&f_{\epsilon}(t)= \exp \left (-\frac{t}{\tau_{\epsilon}} \right ).
\end{eqnarray}
%

%Equation
%nucleon transfer
%Langevin change
%rotation
%T-dependence

%two-body region and one-body region
%nucleon transfer
%Temperature-dependence

\subsection{Dynamical equations}

We then perform trajectory calculations on the time-dependent
unified potential energy using the Langevin equation \cite{arit04,zagr05,zagr07a}

The nucleon transfer for slightly separated nuclei is important in surrogate reactions.
Such intermediate nucleon exchange plays an important role in fusion process at
incident energies near and below the Coulomb barrier as well.
We treat the nucleon transfer using the procedure described
in reference \cite{zagr05,zagr07a};
\begin{equation}
\frac{d \alpha}{dt}=\frac{2}{A_{CN}} D_{A}^{(1)}(\alpha) + \frac{2}{A_{CN}} \sqrt{ D^{(2)}_{A}(\alpha)\Gamma_{\alpha}(t), }
\label{trans1}
\end{equation}
This is a Langevin equation neglecting inertia mass for the mass asymmetry parameter.
It expresses change of the asymmetry parameter $\alpha$ due to drift (first term on the
right hand side) and diffusion (second term on the r.h.s.) processes.
It is obtained by a certain approximation starting from the Master equation for transition
of different particle-hole states.  Such Master equation giving
discrete change of nucleon numbers is transformed to an equation for a continuous variable
$\alpha$ via Fokker-Planck equation to the Langevin equation shown above\cite{zagr05,zagr07a}.

After the window of the touching nuclei opens sufficiently (hereafter ``the mono-nucleus state''), the treatment of the evolution of the mass-asymmetric
parameter $\alpha$ switches from eq.~(\ref{trans1}) to the Langevin equations with the procedure described in reference \cite{arit04}.
The multidimensional Langevin equations \cite{arit04,zagr05,arit09} are now unified as
\begin{eqnarray}
\frac{dq_{i}}{dt}&=&\left(m^{-1}\right)_{ij}p_{j}, \nonumber \\
\frac{dp_{i}}{dt}&=&-\frac{\partial V}{\partial q_{i}}
                 -\frac{1}{2}\frac{\partial}{\partial q_{i}}
                   \left(m^{-1}\right)_{jk}p_{j}p_{k}
                  -\gamma_{ij}\left(m^{-1}\right)_{jk}p_{k} \nonumber \\
                  &&+g_{ij}R_{j}(t), \nonumber \\\frac{d \theta}{dt}&=&\frac{\ell}{\mu_{R}R^{2}}, \nonumber \\
\frac{d \varphi_{1}}{dt}&=&\frac{L_{1}}{\Im_{1}}, \nonumber \\
\frac{d \varphi_{2}}{dt}&=&\frac{L_{2}}{\Im_{2}}, \nonumber \\
\frac{d \ell}{dt}&=&-\frac{\partial V}{\partial \theta}-\gamma_{tan}\left( \frac{\ell}{\mu_{R}R}-\frac{L_{1}}{\Im_{1}}a_{1}-\frac{L_{2}}{\Im_{2}}a_{2}\right)R \nonumber \\
                  &&+Rg_{tan}R_{tan}(t), \nonumber \\
\frac{d L_{1}}{dt}&=&-\frac{\partial V}{\partial \varphi_{1}}+\gamma_{tan}\left( \frac{\ell}{\mu_{R}R}-\frac{L_{1}}{\Im_{1}}a_{1}-\frac{L_{2}}{\Im_{2}}a_{2}\right)a_{1} \nonumber \\
&&-a_{1}g_{tan}R_{tan}(t), \nonumber \\
\frac{d L_{2}}{dt}&=&-\frac{\partial V}{\partial \varphi_{2}}+\gamma_{tan}\left( \frac{\ell}{\mu_{R}R}-\frac{L_{1}}{\Im_{1}}a_{1}-\frac{L_{2}}{\Im_{2}}a_{2}\right)a_{2} \nonumber  \\
&&-a_{2}g_{tan}R_{tan}(t),
%\label{Lan}
\end{eqnarray}
where a summation over repeated indices is assumed.
The collective coordinates $q_{i}$ stands for $z$, $\delta$ and $\alpha$.
The symbol $p_{i}$ denotes momentum conjugate to $q_{i}$, and  $V$ is the multi-dimensional potential energy.
Definition of other parameters is given in Fig.~2:
The symbols $\theta$ and $\ell$ are the relative orientation of
nuclei and relative angular momentum, respectively, and $\varphi_{1}$ and $\varphi_{2}$ denote
the angles of rotation of the nuclei in the reaction plane
(their moments of inertia and angular momenta are $\Im_{1,2}$ and $L_{1,2}$, respectively),
$a_{1,2}=R/2 \pm (R_{1}-R_{2})/2$ are the distances from the centers of the
fragments up to the middle point between nuclear surfaces, and $R_{1,2}$ are the nuclear radii.
The symbol $R$ is the distance between the nuclear centers.
The total angular momentum $L=\ell+L_{1}+L_{2}$ is conserved.
The symbol $\mu_{R}$ denotes the reduced mass, and $\gamma_{tan}$
is the friction force in the tangential direction of colliding nuclei, here we call it as the sliding friction.

%-----------------------------------------------------------------------
%We are interested mainly in the reaction at near-barrier energies.
%The rotation of heavy nuclei is rather slow and orientation effects are remarkable.
%The orientation effects are affected by the initial orientations of statically deformed nuclei.
%Therefore, in reference \cite{zagr05,zagr07a}, equations~(4) were solved numerically assuming $\frac{\partial V}{\partial \varphi_{1}}=\frac{\partial V}{\partial \varphi_{2}}=0$.
%However, in this study,  we assume the nose-to-nose collision as a first approximation.
%The detail is explained in Ref. \cite{zagr05}.
%However, in this study, we assume the nose-to-nose collision as a first approximation.
%-----(Referee)--------------------
%We restrict ourselves to a nose-to-nose geometry of the colliding nuclei, i.e. we take the line connecting the centers of mass of the two colliding nuclei as the common symmetry axis for their deformation.
%-----------<Check1>-------------------

The symbols $m_{ij}$ and $\gamma_{ij}$ stand for elements of
the shape-dependent collective inertia and friction
tensors, respectively.
For separated nuclei, we use the reduced mass and the phenomenological friction
forces with the Woods-Saxon radial form factor as described
in reference \cite{zagr05,zagr07a}.
We switch the phenomenological friction to the friction for mono-nuclear system
using a smoothing function \cite{zagr05,zagr07a}.
For the mono-nuclear system,
the wall-and-window one-body dissipation is adopted for the
friction tensor, and
a hydrodynamical inertia tensor is adopted
in the Werner-Wheeler approximation for the velocity field \cite{bloc78,nix84,feld87}.
The normalized
random force $R_{i}(t)$ is assumed to be of white noise, {\it i.e.},
$\langle R_{i}(t) \rangle$=0 and $\langle R_{i}(t_{1})R_{j}(t_{2})
\rangle = 2 \delta_{ij}\delta(t_{1}-t_{2})$.  Strength of the
random force $g_{ij}$ is given by Einstein relation; $\gamma_{ij}T=\sum_{k}
g_{ij}g_{jk}$, where $T$ is the temperature of the compound
nucleus calculated from the intrinsic energy of the composite
system.

The adiabatic potential energy is defined as
\begin{equation}
V_{adiab}(q,L,T)=V_{LD}(q)+\frac{\hbar^{2}L(L+1)}{2I(q)}+V_{SH}(q,T),
 \label{vt1}
\end{equation}
\begin{equation}
V_{LD}(q)=E_{S}(q)+E_{C}(q),
% \label{Xev}
\end{equation}
\begin{equation}
V_{SH}(q,T)=E_{shell}^{0}(q)\Phi (T),
% \label{Xev}
\end{equation}
\begin{equation}
\Phi (T)=\exp \left(-\frac{E^{*}}{E_{d}} \right),
% \label{Xev}
\end{equation}
where $I(q)$ stands for the moment of inertia of a rigid body with deformation
$q$, $V_{LD}$ and $V_{SH}$ are the potential energy of the
finite-range liquid drop model and the shell correction energy
taking into account the temperature dependence, respectively.
The symbol $E_{shell}^{0}$ denotes
the shell correction energy at $T=0$. The temperature dependent
factor $\Phi (T)$ is discussed in reference \cite{arit06}, where
$E^{*}$ denotes the excitation energy of the compound nucleus.
The shell damping energy $E_{d}$ is chosen as 20 MeV, which is given
by Ignatyuk et al. \cite{igna75}.

The symbols $E_{S}$ and $E_{C}$ denote a generalized surface energy \cite{krap79} and
Coulomb energy, respectively. The centrifugal energy arising from
the angular momentum $L$ of the rigid body is also considered.
%The detail is explained in reference \cite{arit04}.
The intrinsic energy
of the composite system $E_{int}$ is calculated for each trajectory
as
\begin{equation}
E_{int}=E^{*}-\frac{1}{2}\left(m^{-1}\right)_{ij}p_{i}p_{j}-V(q,L,T).
\end{equation}
Here, $E^{*}$ is given by $E^{*}=E_{cm}-Q$, where $Q$ and $E_{cm}$ denote
the $Q$-value of the reaction and the incident energy in the
center-of-mass frame, respectively.
Each trajectory starts from a sufficiently large distance between both
nuclei \cite{arit09}.

%cross section

%The capture and fusion cross sections
%are calculated as follows,

%\begin{eqnarray}
%&&\sigma_{cap}=\frac{\pi\hbar^{2}}{2\mu_{0}E_{cm}}\sum_{\ell=0}^{\infty}(2\ell+1)T_{\ell}(E_{cm},\ell), \\
%&&\sigma_{fus}=\frac{\pi\hbar^{2}}{2\mu_{0}E_{cm}}\sum_{\ell=0}^{\infty}(2\ell+1)P_{CN}(E_{cm},\ell),
%\end{eqnarray}
%where $\mu_{0}$ denotes the reduced mass in the entrance channel. $T_{\ell}(E_{cm},\ell)$ is the capture probability
%of the $\ell$-th partial wave. $P_{CN}(E_{cm},\ell)$ is the probability of forming a compound nuclei in competition with
%quasi-fission events. The method used to estimate the probabilities $T_{\ell}(E_{cm},\ell)$ and $P_{CN}(E_{cm},\ell)$ by a dynamical calculation is explained in the next section.

%---------------------------------

%-------(2010.6.11)-------------------------------------

\subsection{Computation}

Due to difference of the initial impact parameters (or the different initial relative angular momenta),
various kind of reaction processes can occur. Moreover, even the trajectories start with the same
initial impact parameter, reactions proceed in quite different ways
due to the random force in the dynamical
equation, which finally leads to different reaction channels.
By choosing various impact parameters randomly and give proper weights,
whole processes of reactions are
described by the present model; the elastic and inelastic scattering,
deep inelastic collision,
quasi-fission, fusion-fission process, and a few nucleon transfer process (the surrogate
reaction).
They are treated simultaneously by the model.
This is a big advantage of the present approach, since these reactions correlate, thus they
can give information to each other.
An example is a determination of the unknown
parameter $\gamma_{tan}$ through FFMD of fusion-fission like
process as will be discussed below.

\section{Results and discussion}

The unified model, on which the present model is based,
has been applied to several types of reactions
and succeeded to describe the experimental data
\cite{zagr05,zagr07a,zagrt,arit09}.
In our previous study, we precisely investigated the incident energy dependence of
mass distribution of fission fragments in the reactions
$^{36}$S+$^{238}$U and $^{30}$Si+$^{238}$U \cite{arit09,nish08,nish10b}.
The calculation results reproduced the experimental data well and clarified the origin of the fine structure
of the mass distribution of fission fragments at the low incident energy.

Here, we apply the present model to the surrogate reaction.
To evaluate and clarify the model, we focus on the mass distribution of the fission
fragments and compare calculated results with
experimental data.
Furthermore, we investigate spin distributions of the compound nucleus
populated by transfer
reactions and discuss the validity condition of the SRM\cite{chiba10}.
We choose a system of $^{18}$O+$^{A}$U $\rightarrow ^{16}$O+$^{A+2*}$U reaction
where $A=$236 or/and 238.

In the Langevin calculation, the sliding friction is mainly responsible for the dissipation of the
angular momentum \cite{bass80,brom84}, though its value is uncertain.
In the present work, we treat the sliding friction as a parameter of the model and investigate
dependence of the calculation upon this parameter.
Measured FFMD data in the reaction
$^{18}$O+$^{238}$U at $E_{c.m} = 133.5$ MeV is shown by dots in Fig.~\ref{1-3dim3}.
It includes all the fission fragments occurring in the above reaction.
The experimental set-up and the data analysis are nearly the same as the references \cite{nish08,nish10}.
In the experiment, both fission fragments were detected in coincidence by using position-sensitive multiwire
proportional counters (MWPCs). The difference of the setup from the references \cite{nish08,nish10} was the
angles of the detector positions that MWPC1 and MWPC2 were located at $-90^{o}$ and $+90^{o}$ with respect
to the beam direction.

In Fig.~\ref{1-3dim3}, calculated data with $\gamma_{tan} = 1, 5, 10$ and 20 $ \times 10^{-22}$ MeV s fm$^{-2}$ are
denoted by the black, red, green and dark blue histograms, respectively.
It is clearly seen that
the result with $\gamma_{tan} = 5$ reproduces the experimental data very well.
With larger sliding friction, the variance of results becomes smaller.
%It means that more nucleons transfer with larger friction.
We can fix values of unknown parameters in this way.

Then, calculations for the 2 neutron transfer reaction are carried out without
any adjustable parameters.  From such calculations, we can determine spin distributions
of the surrogate reactions $^{18}$O+$^{236}$U $\to$ $^{16}$O+$^{238*}$U and
$^{18}$O+$^{238}$U $\to$ $^{16}$O+$^{240*}$U, and verify if the
2 assumptions proposed by Chiba and Iwamoto\cite{chiba10}, namely,
(1) there exist two surrogate reactions whose spin-parity
distributions of decaying nuclei are almost equivalent, and (2) difference of representative spin values
between the neutron-induced and surrogate reactions is not much larger than 10~$\hbar$,
are really satisfied or not.
%In conjunction with the weak Weisskopf-Ewing
%condition, they form a set of sufficient condition for the SRM to yield
%correct neutron cross sections\cite{chiba10}.

Figure \ref{1-3dim4} shows calculated
spin distributions of compound nucleus $^{240}$U populated
in the reaction
$^{18}$O+$^{238}$U $\rightarrow ^{16}$O+$^{240}$U at an incident energy of $E_{c.m.}=160$ MeV, which
is planned to be carried out at JAEA. Results with various values of the sliding friction
are shown.
We can see that majority of the spin of compound nucleus is much
less than 10~$\hbar$ for each value of the sliding friction although they
diverge depending on $\gamma_{tan}$.
Therefore, it is important for the model to have a
capability to determine values of unknown parameters as shown above.
Our model is particularly powerful since this parameter is determined by
using observables corresponding to other reaction channels, which can be
treated simultaneously with the surrogate reactions of our interest.
Figure \ref{1-3dim5} shows spin distributions of compound nuclei $^{240}$U and $^{238}$U in the transfer reactions
$^{18}$O+$^{238}$U $\rightarrow ^{16}$O+$^{240}$U and $^{18}$O+$^{236}$U $\rightarrow ^{16}$O+$^{238}$U,
respectively with the sliding friction  $\gamma_{tan}=5 \times 10^{-22}$
[MeV s fm$^{-2}]$.  These distributions should be interpreted as a semi-classical estimate of a spin distribution
corresponding to excitation of rotational motion due to
angular momentum transfer occurring in these reactions.
It is easily noticed that the spin distributions of the
compound nuclei populated by the two reactions are almost equivalent.
These results suggest that the assumptions (1) and (2) shown above
for the SRM to work\cite{chiba10} are proved to be correct within this model.
In conjunction with the weak Weisskopf-Ewing condition proposed in
Ref.~\cite{chiba10} (which can be
verified by Hauser-Feshbach theory), the present result suggests that
$^{18}$O+$^{238}$U $\rightarrow ^{16}$O+$^{240}$U and $^{18}$O+$^{236}$U $\rightarrow ^{16}$O+$^{238}$U
reactions can be employed as a pair in the SRM.

%The time evolution of the reaction process is simulated.

We propose that this model as a powerful and useful tool
to describe the surrogate reaction process, even though
it is a semi-classical model.
To describe the transfer reaction process more accurately, we may
have to consider quantum effects precisely.
Quantum mechanical models such as DWBA or CDCC \cite{kami86} would be more appropriate to
describe the nucleon-transfer part of the surrogate reactions.
It will be absolutely necessary if we use light-ion projectiles.
We can use these sophisticated models if they are available and connect the populated
spin-distribution to the later part of the present model.
However, it is difficult to treat
transition probabilities to continuous levels quantum mechanically
as realized in surrogate reactions.
As the first step, therefore,
we try to understand the gross feature of the surrogate
reaction and analyze the reaction mechanism using the present model in this paper.
Especially, it is known that
the dynamical model is useful to discuss the
mass distribution of fission fragments \cite{karp01,asan04}, an important observable of
the surrogate reactions which contains information on the populated
compound nuclei such as the spin distribution\cite{aritomo10a}.

\section{Summary}

We propose a first version of a unified dynamical theory to describe
 the whole process of surrogate reactions; the nucleon transfer,
thermalization and
the decay of the populated compound nuclei.
To realize it, we introduced new procedures to the unified theory
of Zagrebaev and Greiner,
namely, switching of the potential energy surfaces
having very different mass numbers,
Langevin equations depending on
different stages of the reaction and a temperature-dependent
shell correction energy.
Trajectory calculations in
terms of the  Langevin equations are employed on a time-dependent
potential energy surface corresponding to different stages of the surrogate reactions.
After the transfer process,
decay of the populated compound nucleus is calculated
with the initial condition obtained from the preceding transfer process.
This model can yield many observables which can be
compared with experimental data directly.

As an example of the application of the present model to surrogate reactions,
we considered a two nucleon transfer
reaction; $^{18}$O+$^{238}$U $\rightarrow ^{16}$O+$^{240}$U, which is planned to be
performed at JAEA.
We treated the sliding friction as a parameter of the model and discussed
the dependence of the calculation results upon the sliding friction.
Then, we discussed the validity condition of the surrogate
ratio method (SRM). We calculated the spin distribution of the compound nuclei with
several sliding frictions for the compound nucleus $^{240}$U in the reaction
$^{18}$O+$^{238}$U $\rightarrow ^{16}$O+$^{240}$U at the incident energy of $E_{c.m.}=160$ MeV.
The calculation results showed that
the spin of compound nucleus was less than 10$\hbar$ for each value of the sliding friction.
Finally, we discussed spin distributions of
compound nucleus $^{240}$U and $^{238}$U in the transfer reactions
$^{18}$O+$^{238}$U $\rightarrow ^{16}$O+$^{240}$U and $^{18}$O+$^{236}$U $\rightarrow ^{16}$O+$^{238}$U,
respectively.
It was found that the spin distributions of decaying nuclei
populated by the two reactions are almost equivalent.
Therefore it is concluded that if these reactions are used as a pair in the SRM, they
would yield the correct neutron cross sections\cite{chiba10}.
These calculation results suggested validity of the SRM within this model.

In the present model, however,
nuclei were treated as nuclear matter,
and a semi-classical approach was employed
except for the fact that
 we took into account the shell correction energy on the potential energy surface.
Such semi-classical model may be too simple, and we may have to consider
quantum effects in order to describe the reactions more accurately.
Nevertheless, the present model is flexible enough to take account of results of more elaborated models.
We therefore consider that the present model is capable enough, as a first step, to
understood gross features of the surrogate reactions which itself is already quite complicated.

As further studies, we improve the model in the nucleon transfer part by taking into
account the quantum effect more precisely.
After experiments of the surrogate reaction at JAEA are performed, we can compare model predictions with
experimental data, e.g.,  distributions of emission angle and
energy loss of ejectile,
mass, charge and total kinetic energy distributions
of fission fragments from various exclusive fission channels,
and the model will be upgraded successively.

\section{Acknowledgments}

The authors are grateful to Prof. V.I.~Zagrevaev for helpful suggestions and valuable discussion.
The special thanks are deserved to Dr. S.~Hashimoto for his helpful comments.
The diabatic potential and adiabatic potential were calculated using the NRV code \cite{zagr07}.
A part of the numerical calculations was carried out on SX8 at YITP in Kyoto University.
Present study is the result of
''Development of a Novel Technique for Measurement of Nuclear Data
Influencing the Design of Advanced Fast Reactors" entrusted to Japan Atomic Energy Agency
(JAEA) by the Ministry of Education, Culture, Sports, Science and Technology of Japan (MEXT).

%\newpage

%---fig caption, reference, check

%fig1

%fig2
%\begin{figure}
%\centerline{
%\includegraphics[height=.64\textheight]{Fig2-trag1.EPS}}
%  \caption{(Color online) Calculated trajectories with the Langevin equation for the various impact parameters, in the reaction
%$^{18}$O+$^{238}$U $\rightarrow ^{16}$O+$^{240}$U.}
%\label{1-3dim}
%\end{figure}

%fig2
\begin{figure}
\centerline{
\includegraphics[height=.30\textheight]{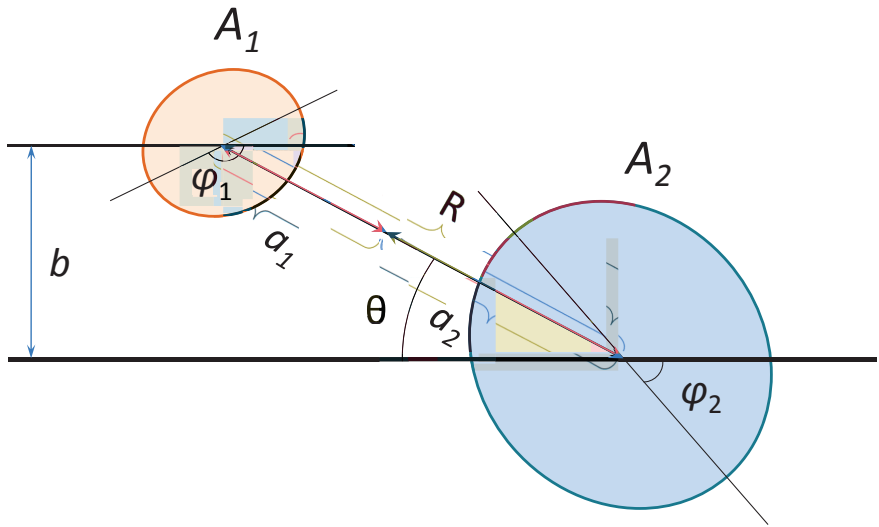}}
  \caption{(Color online) Definition of parameters used in the model.}
\label{1-3dim2}
\end{figure}

%fig12
\begin{figure}
\centerline{
\includegraphics[height=.70\textheight]{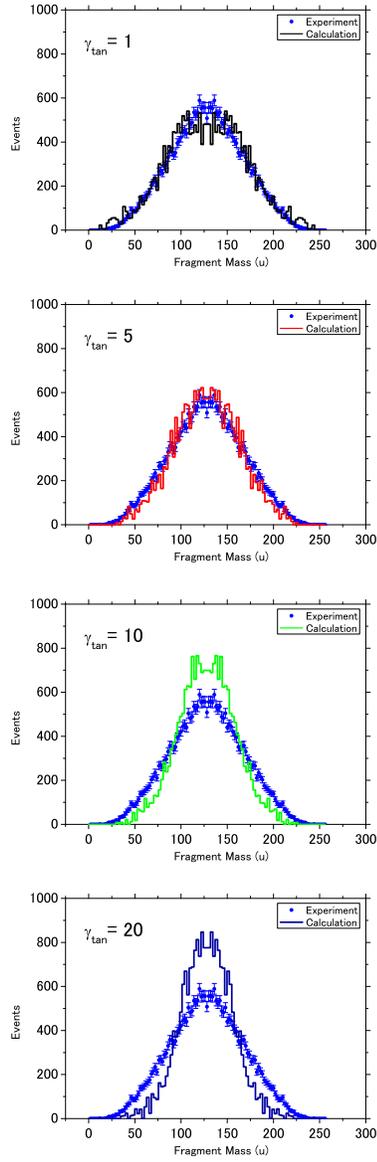}}
  \caption{(Color online) Fragment mass distribution
obtained in the reaction
$^{18}$O+$^{238}$U
at an incident energy of $E_{c.m.}=133.5 $ MeV.
Experimental data and calculation results are denoted by circles and histograms, respectively.
Calculations are shown with sliding frictions $\gamma_{tan}=1, 5, 10$ and $20 \times 10^{-22}$ [MeV s fm$^{-2}]$,
which are multiplied by the factor such that the total cross section agree with the experimental
value to compare the shape of the mass distribution with the experiment.
}
\label{1-3dim3}
\end{figure}

%fig10
\begin{figure}
\centerline{
\includegraphics[height=.44\textheight]{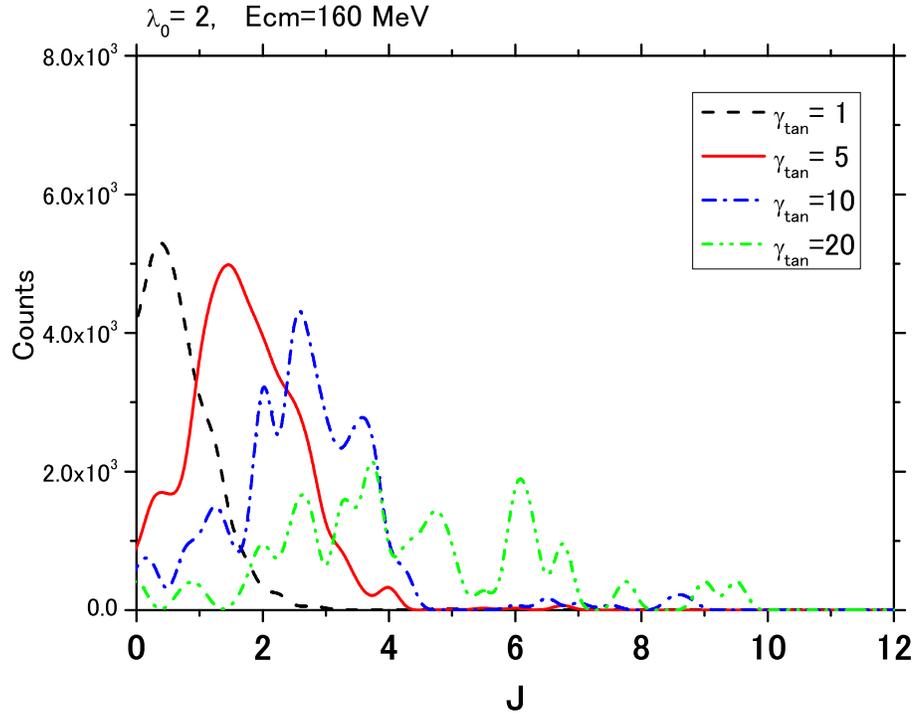}}
  \caption{(Color online) Spin distribution of compound nucleus $^{240}$U in the reaction
$^{18}$O+$^{238}$U $\rightarrow ^{16}$O+$^{240}$U at the incident energy of $E_{c.m.}=160$ MeV
for several sliding frictions.
The black, red, blue and green lines denote for $\gamma_{tan}=1, 5, 10$ and 20
$ \times 10^{-22}$ [MeV s fm$^{-2}]$, respectively..}
\label{1-3dim4}
\end{figure}

%fig11
\begin{figure}
\centerline{
\includegraphics[height=.44\textheight]{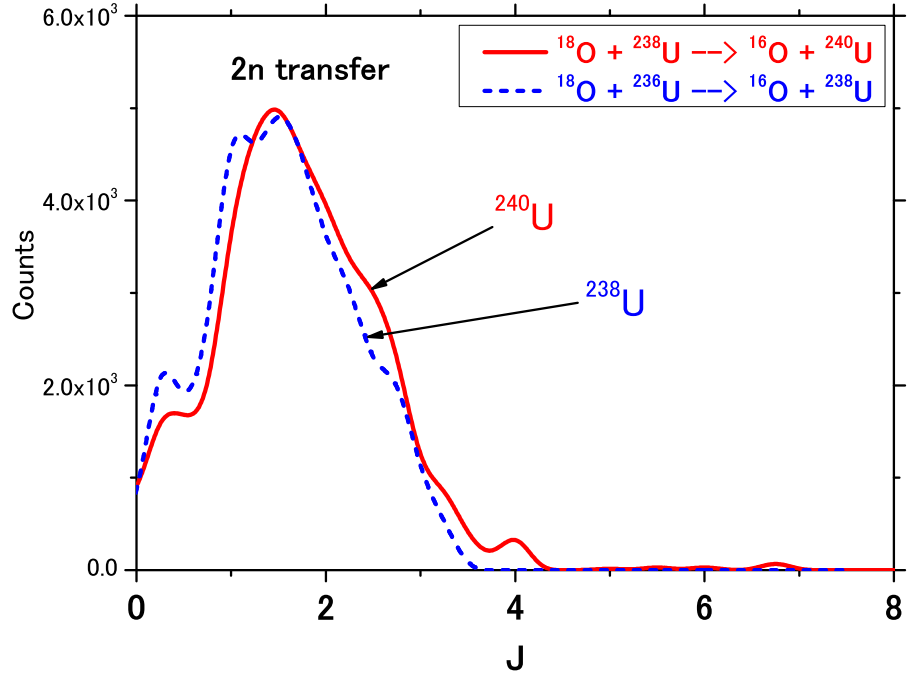}}
  \caption{(Color online) Spin distribution of compound nuclei $^{240}$U and $^{238}$U in the transfer reactions
$^{18}$O+$^{238}$U $\rightarrow ^{16}$O+$^{240}$U and $^{18}$O+$^{236}$U $\rightarrow ^{16}$O+$^{238}$U.,
at the incident energy of $E_{c.m.}=160$ MeV, respectively.
Sliding friction  $\gamma_{tan}=5 \times 10^{-22}$ [MeV s fm$^{-2}]$ is used.}
\label{1-3dim5}
\end{figure}

%-------------------------------------

%\end{document}

\end{document}